\documentclass[12pt,onecolumn]{revtex4}
\usepackage{amsmath}
\usepackage{amssymb,mathrsfs,eufrak}
\usepackage{graphics}
\usepackage{empheq}

\newcommand\ro{\hat\rho}
\newcommand\Ho{\hat H}
\newcommand\Ao{\hat A}

\newcommand\qo{\hat q}
\newcommand\po{\hat p}
\newcommand\phio{\hat\phi}

\newcommand\Mc{\mathcal{M}}

\newcommand\Tr{\mathrm{Tr}}

\newcommand\half{\frac{1}{2}}

\begin{document}
\title{Dissipation in the Caldeira-Leggett model}   
\author{L. Ferialdi}
\email{ferialdi@fmf.uni-lj.si}
\affiliation{Department of physics, Faculty of Mathematics and Physics, University of Ljubljana, Jadranska 19, SI-1000 Ljubljana, Slovenia}
\date{\today}
\begin{abstract}
Caldeira and Leggett (CL) in a seminal paper derived a master equation describing Markovian Quantum Brownian motion. Such an equation suffered of not being completely positive, and many efforts have been made to solve this issue.
We show that, when a careful mathematical analysis is performed, the model considered by CL leads to a non dissipative master equation. We argue that the correct way to understand the master equation derived in the CL regime is to consider it non-Markovian. Moreover, we show that if one wants to provide a microscopic description of Quantum Brownian motion with the CL model, one always needs to consider a non-Markovian dynamics. We conclude that dissipation is a genuinely non-Markovian feature.
\end{abstract}
\maketitle

\section{Introduction}
Quantum Brownian motion and the issue of dissipation in open quantum systems has attracted the attention of physicists for decades. The understanding of dissipative dynamics has recently become urgent, as quantum technologies are approaching regimes where dissipation plays a crucial role.
 This phenomenon has been extensively studied with different approaches~\cite{BrePet02,Wei08}; here we review only the literature on the master equation approach, since it is the one exploited in this paper.
The first analysis of Quantum Brownian motion was phenomenological, and exploited the Lindblad equation  for completely positive Markovian open systems dynamics~\cite{Lin76}. Different authors considered suitable linear combinations of position and momentum operators as generating the Lindblad equation, and analyzed the dissipative dynamics it gave rise to~\cite{Lin76b,SanScu87,Gal93,Gaoeco}. Recently, this phenomenological approach has been generalized to a Lindblad operator that is quadratic in the oscillator position operator~\cite{Lam16}.
 
A first microscopic description of Quantum Brownian motion was provided by Caldeira and Leggett (CL). In their seminal paper, they studied the dissipative dynamics of a harmonic oscillator interacting with a  bosonic bath via a position-position coupling~\cite{CalLeg83} (throughout this paper we will call this the \lq\lq CL model\rq\rq). They derived the well known master equation under some assumptions about the bath structure and temperature. The structure of their master equation is close to, but not precisely, the Lindblad one. Since complete positivity is a crucial requirement for quantum dynamics, efforts were made to derive a completely positive version of the CL master equation, e.g. by relaxing the high temperature condition~\cite{Dio93,Dio95}, or by exploiting the scattering formalism~\cite{LanVac97,Vaceco}. %However, none of them provided a microscopic derivation of such a master equation, starting from the CL model.

Some years later, Hu, Paz and Zhang (HPZ) derived the non-Markovian master equation for Quantum Brownian motion~\cite{HPZ}. Their result generalizes the one by CL, extending it to any thermal bath. Remarkably, their master equation is exact and completely positive. This result, clarified the issue of dissipative dynamics by extending its analysis to the non-Markovian sector.
%Recent results on general Gaussian non-Markovian maps~\cite{DioFer14} and their master equations~\cite{Fer16,Fer17} (that cover also the HPZ case), allowed to have a new perspective on the problem of dissipation in open quantum systems.???
However, the issue of dissipation at the Markovian level is still open: Is it possible to derive a completely positive, dissipative, Markovian dynamics for the CL model? As we will show, the answer to this question is negative. From a broader perspective, the aim of this paper is to clarify the issue of dissipation in the independent oscillators (or bosonic bath) model. We consider the CL model, and we show that if one is careful in taking the Markov limit starting from the approximations introduced by CL, the resulting master equation is actually non dissipative.
We further show that the only way to describe dissipation starting from this model, is by considering a non-Markovian dynamics. We argue that dissipation is a fundamentally non-Markovian feature.

%Accordingly, the studies focused on making the CL master equation in Lindblad form, were biased by a crucial misunderstanding: the CLME is not a Markovian master equation. 

This paper is organized as follows: in Sec. II we introduce the independent oscillators model, and provide exact results on the non-Markovian dynamics related to it. Section III is devoted to the presentation of the CL master equation and the essential steps of its derivation.
In Sec. IV we explain why the CL master equation fails at describing dissipation at an arbitrary time, and how this can be correctly accounted for by a microscopic model. In Sec.V we recall admissible phenomenological descriptions of dissipation, and in Sec.VI we summarize the results of this paper.

\section{The independent oscillators model}
The paradigmatic model exploited to investigate open quantum systems is the so-called \lq\lq independent oscillators model\rq\rq, where the environment is modeled by a set of independent harmonic oscillators~\cite{Foretal88}. We consider an open system whose dynamics is described by the Liouville equation with an Hamiltonian of the following kind: $\Ho=\Ho_S+\Ho_I+\Ho_B$. $\Ho_S$ is the system Hamiltonian, and $\Ho_B$ is the Hamiltonian of the  bath of independent oscillators. The interaction Hamiltonian is assumed to be bilinear, i.e. $\Ho_I=\Ao^j\phio_j$, where $\Ao^j$ are Hermitian system operators, and $\phio_j$ are Hermitian bath fields. We further assume that the initial state of the open system is factorized, i.e. $\ro=\ro_S\otimes\ro_B$, where $\ro_B$ is a Gaussian thermal state completely characterized by its two point correlation function.
 In a recent paper~\cite{DioFer14}, it has been proven that under these conditions the most general completely positive, trace-preserving, non-Markovian Gaussian map reads
\begin{equation}\label{GMt}
\Mc_t=T\exp\left\{\int_0^t\!d\tau\!\!\int_0^t\!ds D_{jk}(\tau,s)\left[\!\Ao^k_{L}(s)\Ao^j_{R}(\tau)\!-\!\theta_{\tau s}\Ao^j_{L}(\tau)\Ao^k_{L}(s)\!-\!\theta_{s\tau}\Ao^k_{R}(s)\Ao^j_{R}(\tau)\!\right]\!\!\right\}\,,
\end{equation}
where $\theta_{\tau s}$ denotes the step function that is $1$ for $\tau>s$, and the bath two-point correlation function is $D_{ij}(t,s)=\Tr_B[\phio_i(t)\phio_j(s)\ro_B]$. The Einstein convention of summing over repeated indexes and the interaction picture are understood. We have also introduced the $L/R$ notation, such that $\Ao^j_L\Ao^k_R\ro=\Ao^j\ro\Ao^k$.
The term \lq\lq Gaussian\rq\rq~refers to the fact that the map is the exponential of quadratic combinations of the system operators $\Ao^j$, and it preserves the Gaussian structure of the initial state.
Although Eq.~\eqref{GMt} represents an important formal characterization of non-Markovian dynamics, when one wants to calculate average values of physical quantities one needs a master equation. This can be achieved when the relevant system Hamiltonian $\Ho_S$ is quadratic if the system is bosonic~\cite{Fer16}, or linear in the case of two-level systems~\cite{Fer17}.
In the former case, the master equation associated to $\Mc_t$ reads: 
\begin{eqnarray}\label{NMME}
\frac{d\ro}{dt}&=&-i[\Ho_S,\ro]+ \Gamma_{jk}(t)[\Ao^j,[\Ao^k,\ro]] +\Theta_{jk}(t) [\Ao^j,[\dot{\Ao}^k,\ro]] \nonumber\\
\label{MEsch}&&+i\Xi_{jk}(t)[\Ao^j,\{\Ao^k,\ro\}]+i\Upsilon_{jk}(t)[\Ao^j,\{\dot{\Ao}^k,\ro\}]\,,
\end{eqnarray}
where the coefficients are determined analytically and have rather complicated expressions, provided in the Appendix. This is the most general non-Markovian Gaussian master equation for non-interacting bosonic systems, it is both trace preserving and completely positive.

An interesting limit case of study is the Markovian one, which is achieved when the environment is memoryless. One can easily check that under this condition, i.e. when the bath is delta-correlated $D_{ij}(t,s)=D_{ij}(t)\delta(t-s)$, the map~\eqref{GMt} reduces to
\begin{equation}\label{GMtL}
\Mc_t=T\exp\left\{\int_0^td\tau D_{jk}(\tau)\left(\Ao^k_{L}(\tau)\Ao^j_{R}(\tau)-\frac{1}{2}\Ao^j_{L}(\tau)\Ao^k_{L}(\tau)-\frac{1}{2}\Ao^j_{R}(\tau)\Ao^k_{R}(\tau)\right)
\right\}.
\end{equation}
which differentiated provides the well known Lindblad equation~\cite{Lin76}:
\begin{equation}\label{lind}
\frac{d\ro}{dt}=D_{jk}(t)\left(\Ao^k_{L}(t)\Ao^j_{R}(t)-\frac{1}{2}\Ao^j_{L}(t)\Ao^k_{L}(t)-\frac{1}{2}\Ao^j_{R}(t)\Ao^k_{R}(t)\right)\ro\,,
\end{equation}
where $D_{ij}$ is a positive definite matrix.
One might argue that a Markovian dynamics might be achieved also for bath correlation functions that are not proportional to a Dirac delta, however later we will show with a simple example that this is possible only for peculiar system-bath couplings.
Before proceeding, it is important to stress a striking difference between the non-Markovian master equation~\eqref{NMME}, and the Lindblad one: the former displays a dependence both on the coupling operators $\Ao^j$ and their time derivative $\dot{\Ao}^j$, while the latter displays the coupling operators only. This difference will play a crucial role for the following considerations on dissipation.

\section{The Caldeira-Leggett master equation}
In their seminal paper, Caldeira and Leggett (CL) considered a harmonic oscillator bilinearly coupled to a bath of harmonic oscillators via the position operators~\cite{CalLeg83}:
\begin{equation}
\Ho= \frac{\po^2}{2m}+\frac{m\omega_S^2}{2}\qo^2+\sum_i \frac{\po_i^2}{2m_i}+\frac{m_i\omega_i^2}{2}(\qo_i-\qo)^2\,,
\end{equation}
where $\qo,\po$ ($\qo_i,\po_i$) are respectively position and momentum operators of the system (bath).
We can rephrase the interaction Hamiltonian of this model in the language of the previous section, as follows:
\begin{equation}\label{HI}
\Ho_I= \Ao \phio
\end{equation}
where $\Ao=\qo$ and $\phio=\sum_i m_i\omega_i^2\qo_i$. With this choice of the bath coupling field one finds that the correlation function is: $D=D^{\mathrm{Re}}+iD^{\mathrm{Im}}$ with
\begin{eqnarray}
\label{Dre}D^\mathrm{Re}(t,s)&=&\hbar\!\int_0^\infty \!\!d\omega J(\omega) \coth\!\left(\frac{\hbar\omega}{2k_B T}\right)\!\cos\omega(t-s),\\
\label{Dim}D^\mathrm{Im}(t,s)&=&-\hbar\!\int_0^\infty \!\!d\omega J(\omega)\sin\omega(t-s)\,,
\end{eqnarray}
where $J(\omega)=\sum_i\frac{m_i\omega_i^3}{2}\delta(\omega_i-\omega)$ is the bath spectral density. The authors assumed an initial thermal state at temperature $T$, and an Ohmic spectral density $J(\omega)=\frac{2m\gamma}{\pi}\omega$ with high frequency cutoff $\Omega$. Exploiting the influence functional formalism~\cite{FeyVer63}, and introducing some approximation (described later), CL derive the following master equation
\begin{equation}\label{CL}
\frac{d\ro}{dt}=-\frac{i}{\hbar}[\Ho_S,\ro]-\frac{2m\gamma k_B T}{\hbar^2}[\qo,[\qo,\ro]]
-\frac{i\gamma}{\hbar} [\qo,\{\po,\ro\}]\,.
\end{equation}
The second term of this equation describes decoherence, while the third one describes friction as it provides damping in the equations for the momentum expectation value $\langle \po\rangle$.
If the CL master equation is recast in the Lindblad form~\eqref{lind}, one finds that the matrix $D_{ij}$ is not positive definite, that implies a non completely positive dynamics. This is a crucial requirement for quantum dynamics, and without satisfying it the CL master equation cannot be considered a valid master equation.
In order to solve this issue, it was proposed to add by hand a \lq\lq minimally invasive\rq\rq term, i.e. a contribution that is negligible under the CL assumptions. The result is the following completely positive master equation~\cite{BrePet02}:
\begin{equation}\label{CCL}
\frac{d\ro}{dt}=-\frac{i}{\hbar}[\Ho_S,\ro]-\frac{2m\gamma k_B T}{\hbar^2}[\qo,[\qo,\ro]] 
 -\frac{i\gamma}{\hbar} [\qo,\{\po,\ro\}]-\frac{\gamma}{8mk_B T}[\po,[\po,\ro]]\,.
\end{equation}
Since the seminal paper by CL, many different derivations of this master equation have been provided in the literature~\cite{SanScu87,Gal93,Dio93,Dio95,LanVac97,Gaoeco,Vaceco}. %However, none of them provided a microscopic description starting from the CL model.
The result by CL is obtained under two important assumptions: \lq\lq high $T$\rq\rq and \lq\lq high $\Omega$\rq\rq, which provide a bath correlation function that is \lq\lq close to\rq\rq~a Dirac delta.
These assumptions and the resulting CL master equation are undoubtedly meaningful under the physical point of view. Indeed, we are not questioning by any means the physical validity of the CL master equation. What we claim, and shortly prove, is that under those assumptions the dynamics is not Markovian. In the independent oscillators model the dynamics is Markovian only when the bath correlation is a Dirac delta, and this happens under the limits $T,\Omega\rightarrow\infty$. Having \lq\lq high $T$\rq\rq and \lq\lq high $\Omega$\rq\rq, under the mathematical point of view is not sufficient to have a delta correlation. Having a correlation function that is  \lq\lq close to\rq\rq~a Dirac delta implies that the environment still has some correlation. This results in some memory that, as small as it is, implies a non-Markovian dynamics.

\section{Dissipation in the CL model}

Let us step back to Section II and apply its results to the CL model. Let us consider the Markovian equation~\eqref{lind}, and replace the coupling operator $\Ao$ with $\qo$ as prescribed by the CL model. What one finds is that the correct Markovian master equation for this model reads
\begin{equation}\label{JZ}
\frac{d\ro}{dt}=-\frac{i}{\hbar}[\Ho_S,\ro]-\frac{2m\gamma k_B T}{\hbar^2}[\qo,[\qo,\ro]]\,.
\end{equation}
This equation is of the Joos-Zeh type~\cite{JoosZeh}, and describes decoherence but not dissipation. 
Actually, one should be surprised that Eq.~\eqref{CL} displays the momentum operator while, as we have previously noticed, the general Markovian dynamics obtained from the interaction with a bosonic bath should display only the coupling operator ($\qo$ in this case). One might now wonder \lq\lq why is the CL result different?\rq\rq. As previously mentioned, it is a consequence of the \lq\lq high $T$\rq\rq~and \lq\lq high $\Omega$\rq\rq~assumptions. If one wants to derive a true Markovian dynamics from CL master equation one needs to take the temperature ($T$) and the cutoff ($\Omega$) to infinity (in order to have a truly memoryless bath). Doing so in Eq.~\eqref{CL}, one sees that the decoherence term diverges, while the dissipative one remains constant, implying that it can be neglected. This is the same reasoning that allowed researchers to add the \lq\lq minimally invasive\rq\rq~term since it was negligible compared to others, and it is quite surprising that they did not choose to neglect the dissipative term (as one should do).
%first two terms are divergent (of the same order as $T\sim\Omega$), while the dissipative one is constant. By renormalizing these divergences, one is left with Eq.~\eqref{JZ}. 
Another way to recover Eq.~\eqref{JZ} starting from Eq.~\eqref{CL} is to observe that when $T\rightarrow\infty$, the correlation function~\eqref{Dre}-\eqref{Dim} reads: $D(t)\sim T\delta(t)+i \delta'(t)$. One then sees that the real part is much bigger than the imaginary one, which is then negligible. Since the dissipative term of~\eqref{CL} derives from a contribution proportional to $D^{Im}$, it can be neglected, leading to~\eqref{JZ}.

The structure of the master equation~\eqref{lind} itself suggests that if one wants to have terms depending on the momentum operator, one needs to generalize the CL model and couple the system to the bath via $\po$. This is the only way to obtain a dissipative term in a Markovian master equation starting from the independent oscillators model. We consider an interaction Hamiltonian of the type $(\qo-\mu\po)\phio$, where $\mu$ is a constant that accounts for the strength of dissipation and $\phio$ has been defined below Eq.~\eqref{HI}. "Mutatis mutandis, this coupling leads to a master equation of the type (10), where the dissipative term is replaced by $\mu [q,[p,\rho]]$. Unfortunately, this term is diffusive and does not describe friction. This is again due to the fact that in the Markov limit the bath correlation function is real, killing the dissipative contributions coming from the anti-commutators." 
The important conclusion we draw is that one cannot describe dissipation at the Markovian level with the CL model.

If one wants to stick to a microscopic description with the independent oscillators model, one needs to disregard the Markovian approximation and consider the exact (non-Markovian) dynamics.
This result was achieved by Hu, Paz and Zhang (HPZ) in their seminal paper~\cite{HPZ}. Their master equation is exact and completely positive: unlike CL, HPZ do not perform any approximation, and the bath correlation function they consider has the general form~\eqref{Dre}-\eqref{Dim}. The HPZ master equation can be obtained by replacing $\Ao=\qo$ in Eq.~\eqref{MEsch}:
\begin{equation}\label{HPZ}
\frac{d\ro}{dt}=-i[\Ho_S-\Xi(t)\qo^2,\ro]+ \Gamma(t)[\qo,[\qo,\ro]] +\Theta(t) [\qo,[\po,\ro]] +i\Upsilon(t)[\qo,\{\po,\ro\}]\,,
\end{equation}
where the coefficients read
\begin{eqnarray}
\label{Gamma}\Gamma(t)&=&-\int_0^tds\,\mathbb{D}^{Re}(t,s)\cos\omega_S(t-s)\\
\label{Theta}\Theta(t)&=&\int_0^tds\,\mathbb{D}^{Re}(t,s)\frac{\sin\omega_S(t-s)}{\omega_S}\\
\label{Xi}\Xi(t)&=&-\int_0^tds\,\mathbb{D}^{Im}(t,s)\cos\omega_S(t-s)\\
\label{Upsilon}\Upsilon(t)&=&\int_0^tds\,\mathbb{D}^{Im}(t,s)\frac{\sin\omega_S(t-s)}{\omega_S}\,,
\end{eqnarray}
The kernel $\mathbb{D}=\mathbb{D}^{Re}+i\mathbb{D}^{Im}$ is a suitable combination of $D^{Re}$ and $D^{Im}$, that in the weak coupling limit simplifies to $\mathbb{D}=D$ (see Appendix).
Since this master equation displays the term $[\qo,\{\po,\ro\}]$ with imaginary coefficient, it correctly describes the dissipative interaction between the system and a thermal bath.
It is useful to rewrite the HPZ master equation in the non-diagonal Lindblad form as follows:
\begin{equation}\label{HPZlin}
\frac{d\ro}{dt}=-i[\Ho_S-\Xi(t)\qo^2-\Upsilon(t)\{\qo,\po\},\ro]+\sum_{i,j} a_{ij}(t)\left(\hat{F}_i\ro\hat{F}_j-\half\left\{\hat{F}_j\hat{F}_i,\ro\right\}\right)
\end{equation}
with $\hat{F}_1=\qo$, $\hat{F}_2=\po$, and
\begin{equation}
a(t)=\left(
\begin{array}{cc}
-2\Gamma(t)&-\Theta(t)+i\Upsilon(t)\\
-\Theta(t)-i\Upsilon(t)&0
\end{array}\right)\,.
\end{equation}
From this expression, one can easily check that one of the two eigenvalues of the matrix $a$ is always negative. This implies that the dynamics described by Eq.~\eqref{HPZ} is non-Markovian for any (non-singular) choice of the bath correlation function $D$. Accordingly, the only way to obtain a Markovian dynamics starting from the HPZ master equation is to choose $D(t,s)\propto\delta(t-s)$, that replaced in Eqs.~\eqref{Gamma}-\eqref{Upsilon} leads to Eq.~\eqref{JZ}, confirming our previous result. Remarkably, this also implies that one can never obtain a time-inhomogeneous Markov dynamics from this model, but only a semigroup dynamics.
If instead we choose the correlation function derived by CL (i.e. $D(t,s)=T\tilde{\delta}(t-s)+i\tilde{\delta}'(t-s)$ where the tilde denotes that these functions are \lq\lq close to\rq\rq~the true deltas), we obtain Eq.~\eqref{HPZ} where the coefficients have a (very) weak dependence on $t$ (see e.g. Eq.(3.399) of~\cite{BrePet02}). We stress that also the microscopic derivation of Eq.~\eqref{CCL} provided in~\cite{Dio93} suffers these same issues.
We then see that the correct way to interpret the master equation derived in the CL regime (or in any other finite temperature/cutoff regime) is to consider it non-Markovian. According to this point of view, Eq.~\eqref{CL} should be modified by adding a term of the type $[\qo,[\po,\ro]]$.

One can further check that if one considers the more general coupling $(\qo-\mu\po)\phio$, the former analysis still holds true, i.e. one can have a Markovian dynamics only with a delta-correlated bath. However, we should mention that if one considers a position-position coupling and performs the rotating wave approximation (typical of quantum optical settings), one obtains a dissipative dynamics also for a bath that is not delta-correlated.

\section{Phenomenological models}
In the previous section we have seen that it is not possible to provide a microscopic description of Markovian Quantum Brownian motion in terms of the CL model.
It is however possible to derive the Markovian master equation~\eqref{CCL} by considering a phenomenological model. The most straightforward approach is to assume that the phenomenology of our system is described by the Lindblad master equation~\eqref{lind} with an operator of the type $\Ao=\qo+i\mu\po$. This is now a legitimate choice since we are considering Eq.~\eqref{lind} as given, and we are no more conditioned by the hermiticity of the system-bath coupling~\cite{SanScu87,Gaoeco,Vaceco}. This option leads to the completely positive master equation~\eqref{CCL}. We should mention however that a choice of this kind is not free of criticism~\cite{Gaoeco}.

Another phenomenological model which is widely used are stochastic Schr\"odinger equations (SSEs)~\cite{SSE}. One assumes that the interaction with the environment can be mimed by a (classical) stochastic process. This results in modifying the Schr\"odinger equation for the wave function by adding a stochastic term. A SSE is said to \lq unravel\rq~ a given master equation when the solution $|\psi_t\rangle$ of the first is such that $\ro_t=\mathbb{E}[|\psi_t\rangle\langle\psi_t|]$ solves the second ($\mathbb{E}$ denoting the stochastic average). One can actually prove that there exist an infinite number of SSEs that unravel the same master equation. In~\cite{DioFer14,Fer16} it has been proven that the most general SSE unraveling the master equation~\eqref{NMME} reads
\begin{equation}\label{SSEdiss}
\frac{d|\psi_t\rangle}{dt}=-i\Ao^j_t
                           \!\left(\!\phi_j(t)\!+\!\int_0^t\!ds 
                                 [D_{\!jk}\!(\!t\!,\!s)\!-\!S_{\!jk}\!(\!t\!,\!s)]\frac{\delta}{\delta\phi_k(s)}\!\right)
\!|\psi_t\rangle\;
\end{equation}
where $\frac{\delta}{\delta\phi_j}$ is a functional derivative, and $\phi_j$ are complex, colored, Gaussian stochastic processes, fully determined by the correlation functions
\begin{eqnarray}
\mathbb{E}\left[\phi_j^\ast(\tau)\phi_k(s)\right]&=&D_{jk}(\tau,s)\,,\\
\mathbb{E}\left[\phi_j(\tau)\phi_k(s)\right]&=&S_{jk}(\tau,s)\,.
\end{eqnarray}
Note that $S$ is not displayed by the gaussian map~\eqref{GMt}: this is where the infinite free parameters of the SSE are encoded.
In the Markovian limit, i.e. when the stochastic process is a white noise, Eq.~\eqref{SSEdiss} simplifies to
\begin{equation}\label{MSSEdiss}
\frac{d|\psi_t\rangle}{dt}=\left(-i\Ao^j_t\phi_j(t)-\half [D_{jk}(t)-S_{jk}(t)]\Ao^j_t\Ao^k_t\right)|\psi_t\rangle\;.
\end{equation}
Also in this case, if we choose $\Ao=\qo+i\mu\po$, Eq.~\eqref{MSSEdiss} unravels Eq.~\eqref{CCL}, proving that this model provides a correct description of Markovian Brownian motion~\cite{BasIppVac05}.

\section{Conclusions}
We have shown that the CL model is not suitable to describe dissipation at an arbitrary time.
One should not be surprised by this behavior. In fact, dissipation is a dynamical feature and one needs to have a dynamical quantity to account for it. This is not the case in the CL model, where the interaction between the system and the environment is \lq\lq static\rq\rq: the system is coupled via the position operator (which is not a dynamical quantity) to a \lq\lq memoryless\rq\rq (Markovian) environment, that cannot keep track of the previous dynamics. 
Accordingly, one should not expect a dissipative dynamics from the CL model, as correctly described by Eq.~\eqref{JZ}. However, what one expects is that the (completely positive version of the) CL equation describes the correct dynamics for time scales that are much longer than the bath memory timescale, when the dynamics can effectively be considered Markovian.

In order to describe dissipation, one needs to have a dynamical interaction. If one wants to have a Markovian description, this can be easily achieved with phenomenological models, e.g. taking a Lindblad operator proportional to $\po$ in~\eqref{lind}, or coupling the system to a (classical) stochastic process via $\po$.
Unfortunately, if one wants to have a microscopic description of dissipation which is still Markovian, coupling a bath with $\po$ is not sufficient as it does not provide the correct damping term.

We then see that the only way to provide a microscopic description of dissipation is by considering a non-Markovian bath, that keeps track of the past interaction with the system. A model of this kind leads to the HPZ master equation that displays the correct dissipative term.

The reason why the CL master equation provides a dissipative dynamics is that it has been obtained by performing approximations which are meaningful under the physical point of view, but are not Markovian from the mathematical side. As we have shown, in order to have a physically memoryless (Markovian) dynamics, it is sufficient to have a \lq\lq high temperature\rq\rq~bath with a \lq\lq high frequency\rq\rq cutoff. However, under the mathematical point of view, the dynamics is Markovian only when temperature and cutoff are taken to infinity. Considering large but finite temperature and frequency cutoff, provides a bath correlation function that is close to (but not exactly) a Dirac delta. This results in a bath that is almost (but not completely) memoryless, i.e. it is correlated. This implies a dynamics which is not Markovian, and this is how the master equation derived in the CL regime should be understood.
The misunderstanding around the CL master equation likely arose because, in analogy to the classical Langevin equation, one expected to have quantum dissipation also at the Markovian level (and this is still true for phenomenological models). 

We think that a microscopic description with position-position coupling (that can be understood as a first order expansion of a generic position potential~\cite{PetVac05}) has a more fundamental nature (see also~\cite{Leg84}). Accordingly, the results of this paper lead us to conclude that dissipation is a purely non-Markovian feature.

\section*{Acknwoledgements}
The author acknowledges fruitful discussions with L. Diosi and G. Gasbarri. This work was supported by the TALENTS$^3$ Fellowship Programme, CUP code J26D15000050009, FP code 1532453001, managed by AREA Science Park through the European Social Fund. 

\section*{Appendix}
The coefficients of the master equation~\eqref{MEsch} provided in~\cite{Fer16} have rather cumbersome expressions. We provide here an improved version of these coefficients. We start by reminding the quadratic system Hamiltonian $\Ho_S$ provides linear Heisenberg equations of motion for the operators $\Ao^j$. The solution of these is linear and always admits to be written as follows:
\begin{equation}
\Ao^j(s)=\mathcal{C}^{j}_k(s-t)\Ao^k(t)+\tilde{\mathcal{C}}^{j}_k(s-t)\dot{\Ao}^k(t)
\end{equation}
where the kernels $\mathcal{C}$, $\tilde{\mathcal{C}}$ explicitly depend on $\Ho_S$. A calculation similar to that performed in~\cite{Fer17} provides the following expressions for the coefficients of~\eqref{NMME}:
\begin{eqnarray}
\Gamma_{jk}(t)&=&-\int_0^tds\,\mathbb{D}^{Re}_{jl}(t,s)\mathcal{C}_{k}^l(s-t)\,,\\
\Theta_{jk}(t)&=&-\int_0^tds\,\mathbb{D}^{Re}_{jl}(t,s)\tilde{\mathcal{C}}_{k}^l(s-t)\,,\\
\Xi_{jk}(t)&=&-\int_0^tds\,\mathbb{D}^{Im}_{jl}(t,s)\mathcal{C}_{k}^l(s-t)\,,\\
\Upsilon_{jk}(t)&=&-\int_0^tds\,\mathbb{D}^{Im}_{jl}(t,s)\tilde{\mathcal{C}}_{k}^l(s-t)\,.
\end{eqnarray}
The kernel $\mathbb{D}=\mathbb{D}^{Re}+i\mathbb{D}^{Im}$ encodes the out-of-equilibrium response of the bath to the system dynamics, and is defined as follows:
\begin{equation}\label{DD}
\mathbb{D}_{ij}(t,s)=\sum_{n=1}^\infty (-1)^{n-1} D^{(n)}_{ij}(t,s)\,,
\end{equation}
where $D^{(1)}\equiv D$, and higher order terms are determined by means of the recursion
\begin{equation}
D^{(n)}_{ij}(t,s)=\int_0^tdt_n\int_0^tds_2\overbracket{\Ao^\alpha(s_2)\Ao^\beta(t_n)}\left[\bar{D}_{\beta j}(t_n,s)D^{(n-1)}_{i\alpha}(t,s_2)+D_{\beta j}(t_n,s)D^{(n-1)*}_{i\alpha}(t,s_2)\right]\nonumber\,,
\end{equation}
with
\begin{equation}
\bar{D}(t_2,s_2)=D^{Re}(t_2,s_2)+iD^{Im}(t_2,s_2)(2\theta_{t_2s_2}-1)\,,
\end{equation}
and
\begin{equation}\label{contr}
\overbracket{\Ao^{\beta}(s_2)\Ao^{\alpha}(t_n)}=\left[\Ao^{\beta}(t_n),\Ao^{\alpha}(s_2)\right]\theta(t_n-s_2)\,.
\end{equation}
We underline that since the system Hamiltonian is quadratic and the Heisenberg equations are linear, the contractions of Eq.~\eqref{contr} are always c-functions.
%We stress that the kernel $\mathbb{D}$ differs from the one obtained in~\cite{Fer17} for the spin-boson model by the definition of $\bar{D}$. 
Equations~\eqref{Gamma}-\eqref{Upsilon} for the HPZ model are easily obtained by replacing $\mathcal{C}(s-t)=\cos\omega_S(s-t)$ and $\tilde{\mathcal{C}}(s-t)=\sin\omega_S(s-t)/\omega_S$ in the equations above.

\end{document}